\documentstyle[aps,prb,eqsecnum,epsf]{revtex}
\begin{document}
\title{Microscopic Functional Integral Theory of Quantum Fluctuations in Double-Layer Quantum Hall 
Ferromagnets}
\author{Yogesh N. Joglekar$^{1,2}$, Allan H. MacDonald$^{1,2}$}
\address{$^1$Department of Physics,\\ University of Texas at Austin,\\ Austin, TX 78705}
\address{$^2$Department of Physics,\\ Indiana University,\\ Bloomington, IN 47405}
\maketitle
\begin{abstract}
We present a microscopic theory of zero-temperature order parameter and pseudospin stiffness 
reduction due to quantum fluctuations in the ground state of double-layer quantum Hall ferromagnets. 
Collective excitations in this systems are properly described only when interactions in both direct 
and exchange particle-hole channels are included. We employ a functional integral approach which is 
able to account for both, and comment on its relation to diagrammatic perturbation theory. We also discuss 
its relation to Gaussian fluctuation approximations based on Hubbard-Stratonovich-transformation 
representations of interactions in ferromagnets and superconductors. We derive remarkably simple 
analytical expressions for the correlation energy, renormalized order parameter and renormalized 
pseudospin stiffness.
\end{abstract}


\section{Introduction}
\label{sec: intro}

Bilayer quantum Hall systems at Landau level filling factor $\nu=1$ have broken symmetry ground 
states that can be regarded either as easy plane ferromagnets or as excitonic superfluids and 
have been extensively studied over the past 
decade~\cite{qhereviews,haf,apb,lb,wenzee,ezawa,sqm,az,km,ky,dns,kynew}.
Bilayer two-dimensional (2D) electron systems consist of a pair of 2D electron gases separated by
a distance $d$ ($d \approx$ 100 \AA) which is comparable to the typical distance between electrons 
within each layer. When a strong magnetic field is applied perpendicular to the electron layers, the 
quantum Hall regime, in which macroscopic Landau level degeneracy quenches the 
kinetic energy and enhances the role of interactions in determining physical properties, is 
reached. Bilayer quantum Hall systems exhibit a rich variety of broken symmetry 
states~\cite{km,ky,dns,kynew} 
depending on Landau level filling factor and on the relative importance of interlayer and 
intralayer Coulomb interactions. For filling factor $\nu=1$, the ground state is completely spin
polarized and spin excitations are gapped because of Zeeman coupling of the electron spins to the
(strong) magnetic field, justifying the neglect of quantum spin dynamics. The only remaining 
degrees of freedom are the intra-Landau level cyclotron orbit centers and the discrete layer 
index which can be treated as a pseudospin label with ``up'' denoting a state localized in the 
upper layer and ``down'' denoting a state localized in the lower layer. At \emph{total} filling 
factor $\nu=1$ this system's ground state has \emph{spontaneous} interlayer phase coherence 
(easy-plane ferromagnetism in the pseudospin language), and is incompressible~\cite{apb,sqm}.
Since each layer has filling factor $\nu=1/2$, a compressible-state filling factor for isolated 
layers, this quantum Hall effect is entirely due to correlations between the two layers. The 
system undergoes a phase transition at a critical layer separation $d_{cr}$ 
from this incompressible QHE state with pseudospin ferromagnetism to a disordered compressible 
state~\cite{apb,sqm}, possibly with other more exotic intervening states~\cite{dns}.

In this paper we present a theory of quantum fluctuations about the Hartree-Fock mean-field-theory 
ordered ground state of the $\nu=1$ bilayer. Our theory is based on an approximate expression for 
the ground state energy which includes quantum fluctuations and generalizes the random phase 
approximation to cases when both direct and exchange interactions are important.  The ground state 
energy itself is of little direct physical relevance, however our approach is sufficiently flexible 
so that we can evaluate the dependence of the ground state energy on external fields and the winding 
wavevector of spiral states, to obtain quantities of physical interest. The most important macroscopic 
parameters characterizing order in quantum Hall bilayer pseudospin ferromagnets are the total pseudospin 
polarization and the pseudospin stiffness $\rho$, which characterizes the energetic cost of long wavelength 
fluctuations of the pseudospin field.  In mean-field theory, all electrons occupy symmetric bilayer 
states, which corresponds to having all pseudospins polarized along the positive $x$-axis.  We will use 
the pseudospin polarization normalized to its mean field value as a dimensionless order parameter, 
$m_x$, which we will determine by evaluating the dependence of ground state energy on interlayer 
tunneling amplitude $\Delta_t$. The pseudospin stiffness will be determined by evaluating 
fluctuation corrections to the energy of spiral states, in which the pseudospin orientation varies
spatially at a constant rate.  

At a formal level, the calculations described in this paper can be adopted to describe the 
influence of quantum fluctuations on the ground state of any easy-plane itinerant electron 
ferromagnet or any superconductor. In the case of a superconductor, the analog of the pseudospin 
stiffness is the superfluid density.  The present work is motivated in part by an interest in 
applications to these more general problems.  The bilayer quantum Hall system is an ideal test case, 
both because its translational invariance permits many elements of the calculations to be performed 
analytically, and because the importance of quantum fluctuations in the ground state can be adjusted 
over a large range simply by adjusting the interlayer spacing. For zero layer separation, the bilayer 
systems is equivalent to a single-layer system with a spin degree of freedom and a Zeeman coupling 
$\Delta_z$ that plays the same role as the tunneling amplitude $\Delta_t$ in spin-polarized bilayer
systems.  In the $d=0$ limit, the 
mean-field ground state is exact\cite{qhereviews} because the interaction term in the Hamiltonian is 
pseudospin invariant. At finite $d$, the difference between the intralayer Coulomb interaction, 
$V_A(\vec{q})=2\pi e^2/\epsilon q$ and the interlayer Coulomb interaction 
$V_E(\vec{q})=V_A(\vec{q})e^{-qd}$ reduces the interaction term's symmetry from SU(2) to U(1). 
Because of this reduced symmetry, only the $\hat z$ component of the total pseudospin polarization 
commutes with the interaction Hamiltonian.
Hence, even though the mean-field-theory splitting between symmetric and antisymmetric
single-particle state energies is enhanced by interactions, a fully pseudospin-polarized state cannot be an 
eigenstate of the microscopic Hamiltonian and in particular, cannot be the ground state.  In the 
description we use, the ground state many-body wavefunction is a linear combination of states with 
collective excitations embedded in the fully-polarized Hartree-Fock mean-field ground state.  

The paper is organized as follows. In section~\ref{sec: affia} we describe the functional integral 
approach to this problem.  It is known that in these systems, both electrostatic \emph{and} exchange 
fluctuations are essential to the collective mode physics~\cite{haf,apb}, with the former dominating 
the cost of $\hat z$-direction pseudospin fluctuations and the latter controlling the easy-plane 
pseudospin stiffness. This situation contrasts with that of metallic ferromagnets in which exchange 
interactions alone determine the collective mode behavior. No standard Hubbard Stratonovich 
(HS) transformation treatment~\cite{negele} can deal with both simultaneously. Section~\ref{sec: affia} 
formally summarizes the generalized Hubbard-Stratonovich we employ that treats direct and exchange channels 
on an equal footing, and establishes some of the notation we will use. Our approach is based on one developed 
previously by Kerman {\it et al.}\cite{kerman} Section~\ref{sec: grpa} formally summarizes the generalized 
random phase approximation (GRPA) for particle-hole correlation functions and its relationship to the 
generalized HS transformation. In section~\ref{sec: adls} we apply our formalism to a double layer system at 
filling factor $\nu=1$. We calculate the dispersion of collective modes, their contribution to the grand
 potential, and the effect of fluctuations on the pseudospin polarization and pseudospin 
stiffness. We conclude the paper with discussion in section~\ref{sec: discussion}.


\section{Auxiliary Field Functional Integral Approach}
\label{sec: affia}
In bilayer quantum Hall systems, due to the difference between interlayer and intralayer Coulomb 
interaction, there is a capacitive energy cost associated with charge-imbalance between the two 
layers. This anisotropy of the Coulomb interaction in pseudospin-space makes it necessary to 
treat the fluctuations in direct and exchange channels on an equal footing. A general way to take 
into account such fluctuations is the Hubbard Stratonovich transformation~\cite{negele}. In this approach, we 
introduce an integral over a Bose field $\phi$ which converts the fermionic two-body interaction in 
the original Hamiltonian into a one-body term coupled with the field $\phi$. The Hamiltonian is 
then quadratic in the fermion operators so that the trace over these degrees of freedom can
be evaluated exactly to obtain an effective action for the Bose field.  
Stationary phase approximations to the  bosonic action yield mean-field theories. 
The nature of the mean-field state, for example 
whether it is characterized by a Hartree mean-field, an exchange mean-field, or a 
pairing mean-field depends upon the way the Hubbard Stratonovich transformation is done.
The nature of the collective excitations that emerge when Gaussian fluctuations are allowed 
in the boson field also change qualitatively.
For example fluctuations in the Hartree mean-field reveal plasmon collective modes, and fluctuations in 
the exchange mean-field of a ferromagnet reveal spin-wave collective modes. 
For a bilayer system, where the collective mode properties are determined by fluctuations in both channels, 
we need a method which treats both Hartree and exchange mean-fields on an equal footing.
It is known that the standard HS transformation 
does \emph{not} yield a Hartree Fock mean-field state and cannot capture \emph{both} Hartree and the 
exchange fluctuations~\cite{negele}.

Kerman, Levit, and Troudet have presented a generalization of the HS transformation which overcomes 
these limitations~\cite{kerman}. To establish our notation (which differs from that of Kerman 
{\it et al.}), we briefly review the previous work which describes an approach for obtaining 
systematic corrections to Hartree-Fock mean-field approximations for the grand potential. Our approach can 
also be used to systematically improve upon Hartree-Fock mean-field approximations for 
the one-particle Green's function~\cite{ynjahm}.

Consider a many-fermion Hamiltonian in second quantized form,
\begin{equation}
\label{eq: affia1}
\hat{H}= 
\sum_{\alpha\beta}K_{\alpha\beta}c^{\dagger}_\alpha c_\beta + 
\frac{1}{2}\sum_{\alpha\beta\gamma\delta}\langle\alpha\beta|V|\gamma\delta\rangle 
c^{\dagger}_\alpha c^{\dagger}_\beta c_\delta c_\gamma,
\end{equation}
where $K$ is a one-body (kinetic) term and $V$ is a two-body interaction. Here $\alpha$ denotes (a 
set of) single-particle quantum numbers. For example, in the case of double layer systems 
$\alpha=(n,k,\sigma)$ where $n$ is the Landau level index, $k$ is the intra-Landau level index and 
$\sigma$ is the pseudospin index. Introducing pair-labels $a=(\alpha'\alpha)$ and the density matrix 
$\hat{\rho}_a=c^{\dagger}_{\alpha'} c_{\alpha}$, the kinetic energy term becomes 
$K_a\hat{\rho}_a$, where summation over repeated indices is assumed. Similarly, the interaction term 
can by written as $V_{ab}\hat{\Omega}_{ab}$ where $\hat{\Omega}_{ab}=:\hat{\rho}_a\hat{\rho}_b:$ is 
the normal-ordered two body interaction and $V_{ab}=\langle\alpha'\beta'|V|\alpha\beta\rangle$.
Then in an obvious matrix notation, Eq. (~\ref{eq: affia1}) becomes 
$\hat{H}=K\hat{\rho}+V\hat{\Omega}/2$. The partition function in the grand canonical ensemble is 
given by
\begin{equation}
\label{eq: affia2}
Z = \mbox{Tr } e^{\alpha\hat{N}-\beta\hat{H}} =\lim_{\epsilon\rightarrow 0} 
\mbox{Tr }e^{\alpha\hat{N}} T\prod^{M}_{j=1}
\left[1-\epsilon K\hat{\rho}_j-\frac{\epsilon}{2}V\hat{\Omega}_j\right],
\end{equation}
where $T$ denotes the (imaginary) time-ordered product, $\hat{N}$ is the number operator, 
$\epsilon=\beta/M$ and we have retained only the imaginary time index. In the thermodynamic limit 
$\alpha/\beta\rightarrow\mu$, the chemical potential. Since the number operator commutes with the 
Hamiltonian we can expand only the $e^{-\beta\hat{H}}$ term. The central idea of this generalized Hubbard 
transformation is to expand around an \emph{arbitrary} two-body interaction $U$, thus treating 
$K\hat{\rho}+U\hat{\Omega}/2$ as the dominant term and $(V-U)\hat{\Omega}/2$ as a perturbation. 
Using the Gaussian identity
\begin{equation}
\label{eq: affia3}
1 =\frac{1}{\sqrt{\det{U}}}
\int\prod_{\gamma\delta}\frac{d\phi_{\gamma\delta}(j)}{\sqrt{2\pi/\epsilon}}
\exp\left[-\frac{\epsilon}{2}\sum_{\alpha\beta\alpha'\beta'}
\phi_{\alpha\beta}(j)U^{-1}_{\alpha\alpha'\beta\beta'}\phi_{\alpha'\beta'}(j)\right]
\end{equation}
at each time index $j$, a typical term in Eq.(~\ref{eq: affia2}) becomes
\begin{equation}
\label{eq: affia4}
\left[1-\epsilon K\hat{\rho}_j-\frac{\epsilon}{2}V\hat{\Omega}_j\right] = 
\int{\cal D}\phi_j e^{-\epsilon\phi_j U^{-1}\phi_j/2}
\left[1-\epsilon K\hat{\rho}_j+\epsilon\phi_j\hat{\rho}_j-\frac{\epsilon^2}{2}V\hat{\Omega}_j 
\frac{\phi_j U^{-1}\phi_j}{{\cal N}^2}\right].
\end{equation}
Here ${\cal N}^2={\cal N}\times{\cal N}$ stands for a repeat sum over the ${\cal N}$ single-particle
state labels in the Hilbert space and ${\cal D}\phi_j$ implies both a product over single-particle labels 
and the relevant
 normalization factors. The term $\phi_j\hat{\rho}_j$ (Bose field coupling to the density operator) 
is odd in $\phi_j$ and does not contribute to the integral. The combination 
$\phi_j U^{-1}\phi_j/{\cal N}^2$ gives a factor of $1/\epsilon$ after integration and we recover 
the interaction term. Thus, even though the interaction term has a prefactor $\epsilon^2$ and the 
kinetic term has a prefactor of $\epsilon$, both terms are of the \emph{same} order. We stress that 
the limit $\epsilon\rightarrow 0$ must be taken \emph{after} all $\phi$ integrals are done. 
Introducing these fields at each time-step $j$ we get
\begin{eqnarray}
\label{eq: affia5}
Z & = & \lim_{\epsilon\rightarrow 0}\int\prod_{j=1}^{M}{\cal D}\phi_j 
\exp\left(-{\cal S}[\phi]\right),  \\
\label{eq: affia6}
{\cal S}[\phi] & = & \frac{\epsilon}{2}\sum_{j=1}^{M}\phi_j U^{-1}\phi_j 
-\ln\mbox{Tr }e^{\alpha\hat{N}} T \prod_{j=1}^{M} 
\left[1-\epsilon K\hat{\rho}_j+\epsilon\phi_j\hat{\rho}_j-\frac{\epsilon^2}{2}V\hat{\Omega}_j 
\frac{\phi_j U^{-1}\phi_j}{{\cal N}^2}\right].
\end{eqnarray}

Systematic approximations to the grand potential are obtained by expanding the bosonic action 
(~\ref{eq: affia6}) around its minimum. In the limit as $\epsilon\rightarrow 0$, the configuration 
of fields $\phi^{0}$ which minimizes the action ($\partial{\cal S}/\partial\phi|_{\phi^{0}} = 0$) is
 given by
\begin{equation}
\label{eq: affia7}
\phi^{0}_{\alpha\beta}(j)=U_{\alpha\alpha'\beta\beta'}
\langle\hat{\rho}_{\alpha'\beta'}(j)\rangle_{\phi^{0}}, \hspace{2cm} j=1,\ldots,M ;
\end{equation}
where we have defined the thermal average as
\begin{equation}
\label{eq: affia8}
\langle\hat{\rho}_{\alpha\beta}(j)\rangle_{\phi^{0}}=
\lim_{\epsilon\rightarrow 0}\frac{\mbox{Tr } e^{\alpha\hat{N}} T \prod\nolimits^{M}_{i=j+1}
(1-\epsilon\hat{h}_i)\hat{\rho}_{\alpha\beta}\prod\nolimits^{j-1}_{i=1}(1-\epsilon\hat{h}_i)}
{\mbox{Tr }e^{\alpha\hat{N}} T \prod\nolimits^{M}_{i=1}(1-\epsilon\hat{h}_i)}
\end{equation}
with the mean-field Hamiltonian $\hat{h}_i=(K-\phi^{0}_i)\hat{\rho}$. We recall that Eq.(~\ref{eq: affia4}) is 
an exact identity and the interaction term with prefactor $\epsilon^2$ is of the same order as the 
kinetic term with the prefactor $\epsilon$ after all the $\phi$-integrals are done. 
Stationary phase approximation, however, corresponds to replacing the measure ${\cal D}\phi_j$ by 
${\cal D}\phi_j\delta(\phi_j-\phi^{0}_j)$. Hence in the stationary phase 
approximation, interaction term with the prefactor $\epsilon^2$ can be neglected and does not contribute to 
the mean-field solution. 

Henceforth let us assume a static mean-field solution so that the mean-field values $\phi^{0}_j$ are
 independent of the time index $j$. It is clear from Eq.(~\ref{eq: affia7}) that the mean-field 
Hamiltonian is solely determined by the trial interaction $U$ and is independent of the actual 
two-body interaction $V$. We get the Hartree (Fock) mean-field by choosing $U=-V$ ($U=+V^{ex}$), 
while $-U=(V-V^{ex})=V^{A}$ (the antisymmetrized interaction) gives the Hartree-Fock mean-field. The 
mean-field grand potential is given by 
\begin{equation}
\label{eq: affia9}
\Omega_0=\frac{1}{\beta}{\cal S}[\phi^{0}]=-\frac{1}{2}\sum_{n,m}f_n f_m U_{nmnm} 
- \frac{1}{\beta}\sum_n\ln(1+e^{\alpha-\beta\epsilon_n}),
\end{equation}
where $\hat{h}|n\rangle = (K-\phi_0)\hat{\rho}|n\rangle=\epsilon_n|n\rangle$ describe the mean-field
 eigenfunctions and eigenenergies, and $f_n = (1+e^{\beta\epsilon_n-\alpha})^{-1}$ are the Fermi 
occupation numbers. The single-particle representation is determined completely by the trial 
interaction $U$ and is independent of the actual two-body interaction $V$. For the mean-field ground
 state energy we get
\begin{equation}
\label{eq: affia10}
\lim_{\beta\rightarrow\infty}\Omega_0\equiv E_0=\sum_h K_{hh}-\frac{1}{2}\sum_{h,h'} U_{hh'hh'}.
\end{equation}
where the subscript $h$ stands for occupied (hole) states. When $U=-V$, we get the direct contribution to 
the ground state energy, whereas only for $U=-V^{A}$ we recover the ground state energy as the expectation 
value of the original Hamiltonian. 

We improve upon the stationary phase approximation by considering quadratic fluctuations around the 
mean field $\phi^{0}$. These, of course, depend upon the true microscopic interaction $V$. Expanding the 
action (~\ref{eq: affia6}) to second order in fluctuating Bose fields $\phi_j=\phi^{0}+\xi_j$ gives
\begin{equation}
\label{eq: affia11}
\frac{\partial^2 {\cal S}}{\partial\phi(j)\partial\phi(j')}=\epsilon M(j,j')
=\epsilon\left[\delta_{jj'}U^{-1}-\epsilon(1-\delta_{jj'})D^{jj'}+\epsilon\delta_{jj'}S\right],
\end{equation}
where only time indices are explicitly shown. The matrices $D$ and $S$ are defined by
\begin{equation}
\label{eq: affia12}
D^{jj'}_{\alpha\beta\gamma\delta}=
\langle\hat{\rho}_{\alpha\beta}(j)\hat{\rho}_{\gamma\delta}(j')\rangle_{\phi^{0}} -
\langle
\hat{\rho}_{\alpha\beta}(j)\rangle_{\phi^{0}}\langle\hat{\rho}_{\gamma\delta}(j')
\rangle_{\phi^{0}}
\end{equation}
and
\begin{equation}
\label{eq: affia13}
S_{\alpha\beta\gamma\delta}=
U^{-1}_{\alpha\gamma\beta\delta}\frac{\langle V\hat{\Omega}\rangle_{\phi^{0}}}{{\cal N}^2} + 
\langle
\hat{\rho}_{\alpha\beta}(j)\rangle_{\phi^{0}}\langle\hat{\rho}_{\gamma\delta}(j)
\rangle_{\phi^{0}}.
\end{equation}
Here $\alpha,\beta$ are the single-particle labels and $j,j'$ stand for the time indices. 
The fluctuation contribution to the grand potential is given by
\begin{eqnarray}
\label{eq: affia14}
e^{-\beta\Omega_2} & = & 
\lim_{\epsilon\rightarrow 0}\int\prod_{j=1}^M {\cal D}\xi_j \exp\left[-\frac{\epsilon}{2}
\sum_{jj'=1}^{M}\xi_j M_{jj'}\xi_{j'}\right],\\
\label{eq: affia15}
& = & \lim_{\epsilon\rightarrow 0}\left[\prod^{M}_{j=1}\frac{1}{\sqrt{\det U}}\right]
\frac{1}{\sqrt{\det M}}.
\end{eqnarray}
We have to be careful about taking the limit $\epsilon\rightarrow 0$ since the dimension of the 
matrix $M$ is of the order $1/\epsilon$ (see Kerman {\it et al.} for details). The resulting grand 
potential can be naturally divided into a quasiparticle contribution and a fluctuation contribution 
from the low-lying collective modes. The quasiparticle contribution is given by
\begin{eqnarray}
\label{eq: affia16}
\Omega_{qp} & \equiv\Omega_0+\frac{1}{2}\mbox{ tr}S &=\frac{1}{\beta}\sum_n \ln(1-f_n) + 
\frac{1}{2}\sum_{n,m}(V^{A}_{nmnm}+ 2 U_{nmnm})f_n f_m, \\
\label{eq: affia17}
E_{qp} &\equiv \lim_{\beta\rightarrow\infty}\Omega_p &=\sum_h K_{hh}+
\frac{1}{2}\sum_{h,h'}V^{A}_{hh'hh'}.
\end{eqnarray}
We emphasize that the quasiparticle contribution reproduces the Hartree-Fock approximation for the 
ground state energy \emph{irrespective} of the choice of trial potential $U$, even though the 
single-particle eigenstates are still determined by $U$. Among all the trial potentials, the 
antisymmetrized interaction $U=-V^{A}$ corresponding to the Hartree-Fock mean-field optimizes
the quasiparticle grand potential, $\partial\Omega_{qp}/\partial U |_{U=-V^{A}}=0$. Hence 
Hartree-Fock mean-field is a particularly good choice for the trial potential. It follows from 
Eq.(~\ref{eq: affia11}) that the fluctuation contribution is given by
\begin{eqnarray}
\Omega_c &= & \frac{1}{2\beta}\mbox{Tr}\ln\left[1-UD\right]-\frac{1}{2}\mbox{tr}(-UD),\nonumber \\
\label{eq: affia18}
 &= & \frac{1}{\beta}\sum_{\omega_\nu>0}\ln\sinh(\beta\omega_\nu/2)-
\frac{1}{\beta}\sum_{\Delta\epsilon_{mn}>0}\ln\sinh(\beta\Delta\epsilon_{mn}/2)+
\frac{1}{2}\sum_{m\neq n}f_m(1-f_n)U_{mnnm}, \\
\label{eq: affia19}
E_c & = & \frac{1}{2}\sum_{\omega_\nu>0}\omega_\nu- 
\frac{1}{2}\sum_{p,h}\left(\epsilon_p-\epsilon_h\right)+\frac{1}{2}\sum_{p,h}U_{hpph},
\end{eqnarray}
where $\Delta\epsilon_{mn}=\epsilon_m-\epsilon_n$ are the mean-field quasiparticle splittings, and 
$\omega_\nu$ are the collective-mode energies obtained by diagonalizing the modified fluctuation 
matrix $\tilde M = U M$. Here, we have used that $\tilde{M}$ is a symplectic matrix whose eigenvalues occur 
in pairs of opposite signs and 
\begin{equation}
\label{eq: affia20}
\det\tilde{M}=\prod_{iu_k}\prod_{\omega_\nu>0}\prod_{\Delta\epsilon_{mn}>0}\left[
\frac{iu_k^2-\omega_\nu^2}{iu_k^2-\Delta\epsilon_{mn}^2}\right],
\end{equation}
where $iu_k=2\pi k/\beta$ is a bosonic Matsubara frequency.
The subscripts $p,h$ in Eq.(~\ref{eq: affia19}) stand for unoccupied (particle) and 
occupied (hole) states, respectively. We can see from Eq.(~\ref{eq: affia19}) that the correlation energy 
is related to the differences between collective mode energies and the Hartree-Fock quasiparticle excitation
 energies.

The functional integral formalism presented here is quite general and allows calculation of grand 
potential and collective modes around \emph{any} mean field. Henceforth we will concentrate on 
expansion around the Hartree-Fock mean field which is accomplished by choosing $U=-V^{A}$. This 
choice gives, at the mean-field level, the exact ground state when the layer separation is zero. 
In the following section we discuss corresponding approximations in diagrammatic perturbation theory.


\section{Generalized Random Phase Approximation}
\label{sec: grpa}
\begin{figure}[h]
\begin{center}
\epsfxsize=4in
\epsffile{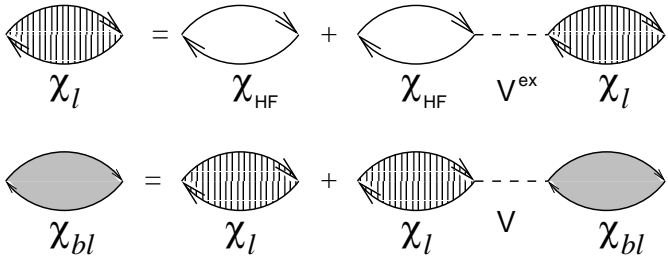}
\vspace{-1cm}
\caption{GRPA for the susceptibility matrix. The ladder-sum susceptibility $\chi_l$ takes into 
account exchange-potential fluctuations. The electrostatic fluctuations which are important in bilayer
systems are taken into account by summing bubbles. $\chi_{bl}$ is the GRPA susceptibility. We have 
written the ladder sum as a bubble sum with an altered interaction matrix element, in keeping with the 
algebraic structure of our formal development in a matrix representation with pair excitation labels.} 
\label{fig: susceptibility}
\end{center}
\end{figure}

In this section we describe the \emph{generalized random phase approximation} (GRPA) for particle-hole  
response functions, which we relate later to the formal fluctuation energy expressions derived in 
the preceding section. Using a general notation, imaginary-time particle-hole response functions 
are defined by 
$\chi^{abcd}(\tau)=\langle T c^{\dagger}_a(\tau)c_b(\tau) c^{\dagger}_c(0) c_d(0)\rangle$. 
For non-interacting electrons these response functions have a simple formal expression
in the representation of single-particle Hamiltonian eigenstates:
\begin{equation}
\label{eq: grpa1}
\chi^{n'nm'm}(iu_k) \equiv D^{n'm'}(iu_k)\delta_{n'm}\delta_{m'n}=(-1)
\left[\frac{f_{n'}-f_{m'}}{iu_k-\Delta\epsilon_{m'n'}}\right]\delta_{n'm}\delta_{m'n},
\end{equation}
where we have Fourier transformed their imaginary time dependence and $iu_k=2\pi k/\beta$ is a bosonic 
Matsubara frequency.  These response functions have poles at the non-interacting electron 
particle-hole excitation energies, 
$\Delta\epsilon_{m'n'}=\epsilon_{m'}-\epsilon_{n'}$~\cite{negele}.
In the \emph{generalized random phase approximation} (GRPA), the single particle eigenstates
are replaced with those obtained by diagonalizing the Hartree-Fock Hamiltonian $\hat{h}_{HF}$ and 
collective fluctuations in direct and exchange potentials are captured by summing ``ladders'' and 
``bubbles'' as summarized diagrammatically in Fig.~\ref{fig: susceptibility}. Including these 
corrections shifts the response function poles from differences of single-particle energies to 
collective excitation energies. Consider, for example, a typical ladder diagram having $p$ 
interaction lines. In a convenient matrix notation, the value of such a diagram is given by
\begin{eqnarray}
\label{eq: grpa2}
\chi^{p}(iu_k) & = & \overbrace{D(iu_k)V^{ex}D(iu_k)\cdots V^{ex}D(iu_k)}^{(2p+1)
\mbox{ terms}}, \\
\label{eq: grpa3}
\langle a b|V^{ex}D|c d\rangle &\equiv &\langle b c |V^{ex}|a d\rangle D^{cd} = 
\langle c b|V|a d\rangle D^{cd}.
\end{eqnarray}
Here the $p^{th}$ order diagram has been written in terms of the Hartree-Fock particle-hole response 
function $D$ and the exchange interaction $V^{ex}$. These diagrams capture the effect of exchange 
fluctuations around the Hartree-Fock mean-field. Adding these contributions for all $p$ gives the 
ladder-sum approximation to the susceptibility. This geometric series can be formally summed to yield
 the following matrix for the ladder-sum response function $\chi_l$
\begin{equation}
\label{eq: grpa4}
\chi_l(iu_k) = D(iu_k)\cdot\left[1-V^{ex}D(iu_k)\right]^{-1}.
\end{equation}
Finally we consider a typical bubble diagram where instead of the Hartree-Fock response function $D$,
the ladder-sum response function $\chi_l$ is used. In matrix notation, a term with $m$ interaction 
lines becomes
\begin{equation}
\label{eq: grpa5}
\chi^{m}(iu_k)=\underbrace{\chi_l(iu_k)V\chi_l(iu_k)\cdots V\chi_l(iu_k)}_{(2m+1) \mbox{ terms}}.
\end{equation}
Again, summing the resulting geometric series we get the following expression for the 
bubble-and-ladder sum susceptibility $\chi_{bl}$ 
\begin{eqnarray}
\label{eq: grpa6}
\chi_{bl}(iu_k)^{-1} = & \chi_l(iu_k)^{-1} +V = & D(iu_k)^{-1} + V^{A}.
\end{eqnarray}
In general, the determinant of the response function matrix $\chi_{bl}$ has isolated poles and 
branch-cuts. By convention, we regard the well-defined collective modes due to the isolated zeros of 
$\det(1+V^{A}D)$ as the collective modes, whereas zeros which lie in a continuum in the thermodynamic limit 
are associated with renormalized particle-hole excitation energies. Assuming that fluctuations here include 
only the collective modes, we get
\begin{equation}
\label{eq: grpa7}
\det\chi_{bl}(iu_k)=\prod_{\omega_\nu>0}\left[\frac{-1}{iu_k^2-\omega_\nu^2}\right]
\end{equation}
where $\omega_\nu$ are the collective-mode frequencies determined by the (isolated) zeros of 
$\det(1 + V^{A}D)$.

With these general expressions it is easy to see the equivalence between the functional integral 
approach and diagrammatics. Let us consider the collective-mode dispersions obtained from the 
response function (diagrammatics) and the quadratic fluctuations (functional integral approach). 
Using antisymmetrized interaction as the tunable potential, $U=-V^{A}$, the determinant of the 
modified fluctuation matrix $\tilde M=UM$ [Eq.(~\ref{eq: affia11})] becomes
\begin{equation}
\label{eq: grpa8}
\det\tilde M = \prod_{iu_k}\det\left[1+V^{A}D(iu_k)\right].
\end{equation}
Comparing (~\ref{eq: grpa6}) and (~\ref{eq: grpa8}), we see that collective-mode energies 
obtained from these two methods are identical. In other words, quadratic fluctuations around a 
Hartree-Fock mean-field give the same result as the GRPA (bubbles and ladders) {\it as far as 
collective-mode dispersion is concerned}. If we had used an exchange-field (Hartree-field) standard 
Hubbard-Stratonovich transformation, the Gaussian fluctuations would have had resonances at the 
poles of $\chi_l$ ($\chi_b$). It is clear that whenever $\chi_{bl}$ has a pole structure that is grossly 
different from that of both $\chi_l$ or $\chi_b$, neither standard HS approach will be adequate. Bilayer 
quantum Hall systems provide one example of such a circumstance.

\begin{figure}[h]
\begin{center}
\epsfxsize=5in
\epsffile{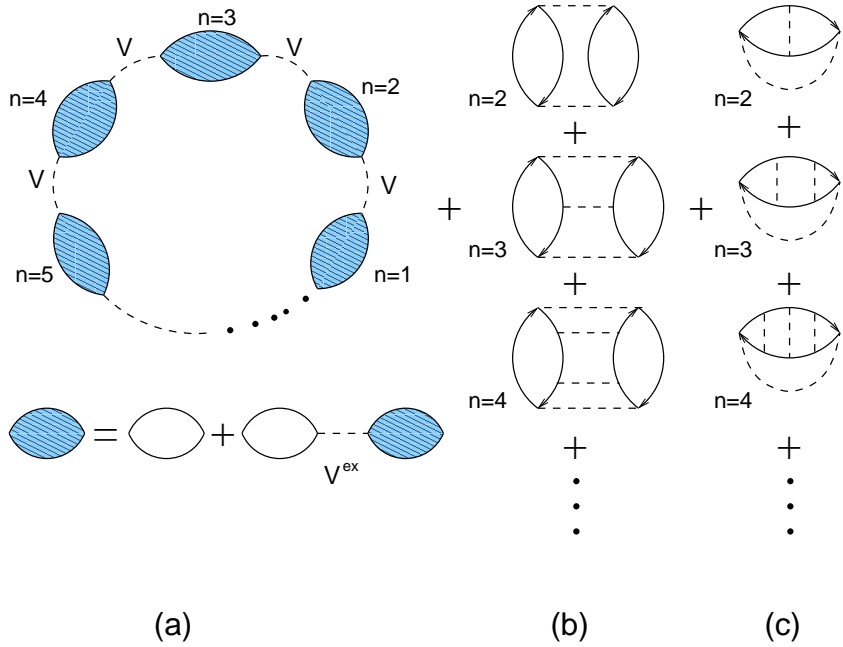}
\vspace{-1cm}
\caption{Diagrammatic content of Eq.(~\ref{eq: grpa10}). The set of diagrams on the left, (a), is a 
random-phase-approximation-like bubble sum, but here the bubbles have ladder diagram vertex corrections. 
Only diagrams with $n\geq 2$ bubbles appear in this set. The symmetry factors of these diagrams match those 
obtained by expanding the $\ln$ as usual. The sets of diagrams on the right, labelled (b) and (c), appear
naturally in the functional integral formalism. All these sets of diagrams can be summed explicitly for 
double layer quantum Hall systems.}
\label{fig: energy_grpa}
\end{center}
\end{figure}

The relationship between the two approaches requires more care when approximating the grand potential. 
The venerable random phase approximation consists of summing the set of bubbles with $n\geq 2$ 
interaction lines and gives the following fluctuation contribution to the grand potential~\cite{negele} 
\begin{equation}
\label{eq: grpa9}
\Omega_b=\frac{1}{2\beta}\mbox{ Tr }\ln\left[1+VD\right]-\frac{1}{2}\mbox{ tr}\left(VD\right).
\end{equation}
This is the same as the grand potential calculated from quadratic fluctuations 
around the Hartree mean-field as can be checked by using $U=-V$ in Eq.(~\ref{eq: affia18}). 
Similarly summing the ``bubbles'' with antisymmetrized interaction $V^{A}$ gives 
\begin{equation}
\label{eq: grpa10}
\Omega=\frac{1}{2\beta}\mbox{ Tr }\ln\left[1+V^AD\right]-\frac{1}{2}\mbox{ tr}\left(V^AD\right).
\end{equation}
In the functional integral approach, this contribution is that obtained by considering quadratic fluctuations 
around the Hartree-Fock mean-field. The diagrammatic content of Eq.(~\ref{eq: grpa10}) is shown in 
Fig.~\ref{fig: energy_grpa}. The set of diagrams on the left, (a), consists of vertex-corrected bubbles with 
$n\geq 2$ interaction lines. These diagrams can be formally summed and we obtain 
$\mbox{Tr}\ln\left[1+V\chi_l\right]-\mbox{tr} (V\chi_l)$. This set of diagrams has been used in the literature to 
approximate the grand-potential when both, direct and exchange, fluctuations are important~\cite{kmoonprl}. 
However the formal expression for the grand potential that follows from the functional integral approach includes 
another class of diagrams which are shown on the right in Fig.~\ref{fig: energy_grpa}. The set of diagrams with 
$n\geq 2$ interaction lines, labelled (b), gives the contribution 
$\mbox{Tr}\ln\left[1-V^{ex}D\right]-\mbox{tr} (-V^{ex}D)$, whereas summing the diagrams in the set labelled (c) 
we get $\mbox{tr} (V\chi_l)-\mbox{tr} (VD)$. The equivalence of sum of these three contributions with the 
formal expression is most easily seen 
by using the fact that $\det\left[1+V^{A}D\right]=\det\left[1+V\chi_l\right]\cdot\det\left[1-V^{ex}D\right]$. In 
identifying the Feynman diagram content of Eq.(~\ref{eq: grpa10}) it is necessary to take care in 
using the pair-state matrix notation that simplified the formal developments in section~\ref{sec: affia}.
 As we will see, the inclusion of this set of diagrams substantially alters the renormalization of the 
system parameters. We refer to the sum all three classes of diagrams as the GRPA.


\section{Application to double layer systems}
\label{sec: adls}

In this section, we apply the formalism developed above to bilayer quantum Hall systems at 
\emph{total} filling factor $\nu=1$.  We neglect spin so that the pseudospins and the orbit centers 
within a Landau level are the only dynamical degrees of freedom. Due to the properties of Landau level 
wavefunctions, there exists a unitary transformation which allows us to associate a 
two-dimensional wavevector with a pair of orbit center labels~\cite{kh}. In other words, momentum 
is a good quantum number for the particle-hole pairs. This property is responsible for the considerable 
progress that can be achieved analytically in the following calculations. We want to calculate the 
correction to the mean-field ground state energy (or grand potential) due to collective excitations 
for spiral states and ferromagnetic states with interlayer coupling. The renormalized 
order parameter $m_x$ and the pseudospin stiffness $\rho$ are obtained from the ground state energy 
by taking appropriate derivatives. 

The Hamiltonian for a double layer system with interlayer tunneling amplitude $\Delta_{t}$ is 
\begin{eqnarray}
\label{eq: adls1}
\hat{H_0}-\mu\hat{N}&=&-\frac{\Delta_t}{2}\sum_{k,\sigma}
c^{\dagger}_{k\sigma}c_{k\overline{\sigma}},\\
\label{eq: adls2}
\hat{V} & = & \frac{1}{2}\sum_{k_i,\sigma_i}
\langle k_1\sigma_1 k_2\sigma_2| V_0 + \sigma_1\sigma_2 V_x|k_3\sigma_1 k_4\sigma_2\rangle 
c^{\dagger}_{k_1\sigma_1}c^{\dagger}_{k_2\sigma_2}c_{k_4\sigma_2}c_{k_3\sigma_1}.
\end{eqnarray}
Here, $\sigma=\uparrow=-\overline{\sigma}$ denotes a state in top layer, while $\sigma=\downarrow$ 
denotes a state in the bottom layer. The $k_i$ the are $y$-components of the canonical momenta which are 
good quantum numbers of the single-particle Hamiltonian in the Landau 
gauge ${\bf A}=(0,Bx,0)$, $V_0=(V_A+V_E)/2$ and $V_x=(V_A-V_E)/2$ are sums and differences of interlayer and 
intralayer Coulomb interactions. $V_x$ characterizes the anisotropy of the interaction in the pseudospin 
space.  The energetic splitting between symmetric and the antisymmetric Hartree-Fock eigenstates is 
enhanced by interactions, $\Delta_{SAS}=\Delta_t+\Delta_{sb}$, where
\begin{eqnarray}
\label{eq: adls3}
\Delta_{sb} & = & \Gamma_E(0), \\
\label{eq: adls4}
\Gamma_{\lambda}(\vec{q})& = &\frac{1}{A}\sum_{\vec{p}}V_{\lambda}(\vec{p})
e^{-p^2l^2/2}e^{i\hat{z}\cdot(\vec{q}\times\vec{p})l^2}
\end{eqnarray}
and $\lambda=0,x,A,E$. The $\Gamma_{\lambda}(\vec{q})$ are interactions between an electron and an 
exchange hole~\cite{kh} whose center is separated by a distance $ql^2$. 

First we concentrate on the case with zero interlayer tunneling. If the external tunneling
is absent, $\Delta_{t}=0$, there is a rotational symmetry in the $x-y$ plane in the 
pseudospin-space. In this case the Hartree-Fock mean-field equations support a class of ``tumbling'' or 
spiral order-parameter solutions $m(x)=(\cos Qx,\sin Qx)$. In a superfluid language, this corresponds to a 
supercurrent solution parameterized by a pairing wave-vector $\vec{Q}=Q\hat{x}$~\cite{ehole}. There are a 
series of metastable spiral ground states with different values of 
$Q$. The $Q$-dependent eigenstates and eigenenergies are given by
\begin{eqnarray}
\label{eq: adls5}
|k,\pm\rangle & = & \frac{1}{\sqrt{2}}\left[|\uparrow\rangle \pm e^{iQkl^2} |\downarrow\rangle\right],
 \\
\label{eq: adls6}
\epsilon^{0}_{-} & = & \Gamma_E(Q)/2  = \Delta_Q/2= -\epsilon^{0}_{+}.
\end{eqnarray}
For $Q=0$ we get the symmetric and the antisymmetric states as eigenstates and the spiral state reduces to 
the ferromagnetic state. By 
evaluating the fluctuation correction to the ground state energy at all values of $Q$ we will be 
able to estimate fluctuation corrections to the spin-stiffness of bilayer quantum Hall ferromagnets. 
The mean-field Green's function is diagonal in the Hartree-Fock eigenstate basis. 
The Hartree-Fock susceptibility $\chi^{+--+}_{HF}=D^{+-}$ is therefore given by 
\begin{equation}
\label{eq: adls7}
D^{+-}(\vec{q},i\Omega) = 
-\frac{e^{-q^2l^2/2}}{2\pi l^2}\frac{\delta n}{(i\Omega_n-\Delta_Q)}=
D^{-+}(\vec{q},-i\Omega),
\end{equation}
where $\delta n=(n_{+}-n_{-})$ is the difference between Fermi factors. As expected the Hartree-Fock
 susceptibility diverges at the quasiparticle energy gap $\Delta_Q$. Due to the absence of dispersion in 
Landau band energies, the ladder-sum for the susceptibility can be evaluated analytically. We find that
\begin{eqnarray}
\label{eq: adls8}
\chi_l^{+--+}(\vec{q},i\Omega) & = & -\frac{e^{-q^2l^2/2}}{2\pi l^2}
\left[
\frac{i\Omega_n+\Delta_Q-\delta n\Gamma_{++}(\vec{q})}{i\Omega^2-{\cal E}^2(\vec{q})}
\right]\delta n,\\
\chi_l^{-++-}(\vec{q},i\Omega) & = & \chi_l^{+--+}(\vec{q},-i\Omega),\nonumber \\
\label{eq: adls9}
\chi_l^{+-+-}(\vec{q},i\Omega) & = & -\frac{e^{-q^2l^2/2}}{2\pi l^2}
\left[\frac{\Gamma_{+-}(\vec{q})\delta n}{i\Omega^2-{\cal E}^2(\vec{q})}\right]\delta n =
\chi_l^{-+-+}(\vec{q},i\Omega).
\end{eqnarray}
Here we have introduced
\begin{eqnarray}
\label{eq: adls10}
\Gamma_{++}(\vec{q})&=&
\left[2\Gamma_A(\vec{q})+\Gamma_E(\vec{q}+Q\hat{x})+\Gamma_E(\vec{q}-Q\hat{x})\right]/4, \\
\label{eq: adls11}
\Gamma_{+-}(\vec{q})&=&
\left[2\Gamma_A(\vec{q})-\Gamma_E(\vec{q}+Q\hat{x})-\Gamma_E(\vec{q}-Q\hat{x})\right]/4, \\
\label{eq: adls12}
{\cal E}^2 &  = & (\Delta_Q-\Gamma_{++})^2-\Gamma_{+-}^2.
\end{eqnarray}
Note that the ladder-sum susceptibility $\chi_l$ has a pole at $i\Omega_n={\cal E}(\vec{q})$. It is 
clear from Eq.(~\ref{eq: adls12}) that this pole arises due to \emph{exchange} fluctuations. In the 
functional integral approach, $i\Omega_n={\cal E}(\vec{q})$ denotes the zeros of 
$\det\left(1-V^{ex}D\right)$. Since all other elements of the susceptibility matrix vanish, we will 
henceforth use only two indices to denote the matrix elements, $\chi^{+--+}\equiv\chi^{+-}$, 
$\chi^{+-+-}\equiv\chi^{++}$ and so on. Furthermore, since 
$\chi^{+-}(\vec{q},i\Omega)=\chi^{-+}(\vec{q},-i\Omega)$ and $\chi^{++}=\chi^{--}$, there are only 
two independent matrix elements.

We include the bubble diagrams to take into account electrostatic effects. This gives the 
following equation for the susceptibility matrix $\chi_{bl}(\vec{q},i\Omega_n)$
\begin{equation}
\label{eq: adls13}
\left[\begin{array}{c}\chi^{+-}_{bl}\\ \chi^{++}_{bl}\end{array}\right]
= \left[\begin{array}{c}\chi^{+-}_{l}\\ \chi^{++}_{l}\end{array}\right]
+ (-2\pi l^2 e^{q^2 l^2/2} v_x^Q)\cdot 
\left[\begin{array}{cc}
\left(\chi^{+-}_l+\chi^{++}_l\right)&\left(\chi^{+-}_l+\chi^{++}_l\right) \\
\left(\chi^{++}_l+\chi^{-+}_l\right)&\left(\chi^{++}_l+\chi^{-+}_l\right) 
\end{array}\right]
\left[\begin{array}{c}\chi^{+-}_{bl}\\\chi^{++}_{bl}\end{array}\right].
\end{equation} 
Here $2\pi l^2 v_x^Q(\vec{q})=e^{-q^2l^2/2}\left[V_A(\vec{q})-V_E(\vec{q})\cos Qq_yl^2\right]/2$ 
represents the electrostatic fluctuation effects. This GRPA susceptibility has a pole at the the 
collective-mode energy given by~\cite{kmoonprl}
\begin{eqnarray}
\label{eq: adls14}
E^2_{sw}(\vec{q},Q) & = & a_Q(\vec{q})\cdot b_Q(\vec{q}), \\
\label{eq: adls15}
a_Q(\vec{q})& = & \Delta_Q-\Gamma_A(\vec{q})+2v_x^Q(\vec{q}), \\
\label{eq: adls16}
b_Q(\vec{q})& = &\Delta_Q-
\frac{1}{2}\left[\Gamma_E(\vec{q}+Q\hat{x})+\Gamma_E(\vec{q}-Q\hat{x})\right].
\end{eqnarray}
In the functional integral approach the same result for collective mode dispersion is obtained from the
zeros of determinant of the fluctuation matrix $\tilde{M}=\left(1+V^{A}D\right)$. 
Since momentum and frequency are good quantum numbers, the fluctuation matrix is diagonal in these indices, 
and effectively only has pseudospin labels. In the eigenstate representation, the interaction matrix 
elements are given by
\begin{eqnarray}
\label{eq: adls17}
\langle k_1\sigma_1,k_2\sigma_2|\hat{V}|k_3\sigma_3,k_4\sigma_4\rangle
& = & 
\frac{1}{4}\langle k_1 k_2 | V_A | k_3 k_4\rangle\left[1+\sigma_1\sigma_2\sigma_3\sigma_4\right] 
\\ \nonumber
& + & 
\frac{1}{4}\langle k_1 k_2 | V_E | k_3 k_4\rangle
\left[\sigma_1\sigma_3 e^{iQ(k_3-k_1)l^2}+\sigma_2\sigma_4 e^{iQ(k_4-k_2)l^2}\right],
\end{eqnarray}
where $k_i$ are the intra-Landau level indices and $\sigma_i=\pm 1$ denote the $k$-dependent 
pseudospin eigenstates. Further constraints on the pseudospin labels are imposed by the fact that the
 Hartree-Fock susceptibility is nonzero only when the particle-hole pseudospins are opposite. 
Hence, the relevant part of the fluctuation matrix becomes
\begin{equation}
\label{eq: adls18}
\langle\sigma_1\sigma_2|1+V^A(\vec{q})D(i\Omega_n)|\sigma_3,\sigma_4\rangle
= 
\left[\begin{array}{cc}
1+ (v_x^Q-\Gamma_{++})\frac{\delta n}{i\Omega_n+\Delta_Q} & 
-(v_x^Q-\Gamma_{+-})\frac{\delta n}{i\Omega_n-\Delta_Q} \\ 
(v_x^Q-\Gamma_{+-})\frac{\delta n}{i\Omega_n+\Delta_Q} & 
1-(v_x^Q-\Gamma_{++})\frac{\delta n}{i\Omega_n-\Delta_Q}
\end{array} \right].
\end{equation}
We emphasize that this $2\times 2$ submatrix of the full $4\times 4$ fluctuation matrix has the same
 determinant as the entire $4\times 4$ matrix. Hence it is sufficient to evaluate this submatrix as far as 
the contribution to the grand 
potential is concerned. In other problems, for example, when considering the effect of collective modes on the
 one-particle Green's function~\cite{ynjahm}, the entire collective-mode propagator is required. 
Notice that in the functional integral approach, the effect of Hartree fluctuations appears via the use of 
antisymmetrized interaction $U=-V^A$ instead of exchange interaction $U=V^{ex}$. Since a spiral state 
spontaneously breaks the rotational symmetry in the $x-y$ plane in pseudospin space, collective 
modes are pseudospin-waves where the pseudospin polarization deviates slowly from the spiral-state 
polarization. The term $a_Q(q)$ in Eq.(~\ref{eq: adls15})
 expresses the energy cost of fluctuations {\it out} of the $x-y$ plane. These fluctuations correspond
 to moving charge from one layer to the other and are finite even at zero wavevector, $a_Q(q=0)\neq 0$. 
In contrast the term $b_Q(q)$ represents the cost of 
pseudospin rotations {in} the $x-y$ plane. In particular the fact that $b_Q(q=0)=0$ indicates the Goldstone 
nature of these collective excitations. We stress that the preceding results for spiral-state susceptibility 
matrix and collective-mode dispersion are valid only when there is no external tunneling between the layers.

\begin{figure}[h]
\begin{center}
\epsfxsize=4in
\epsffile{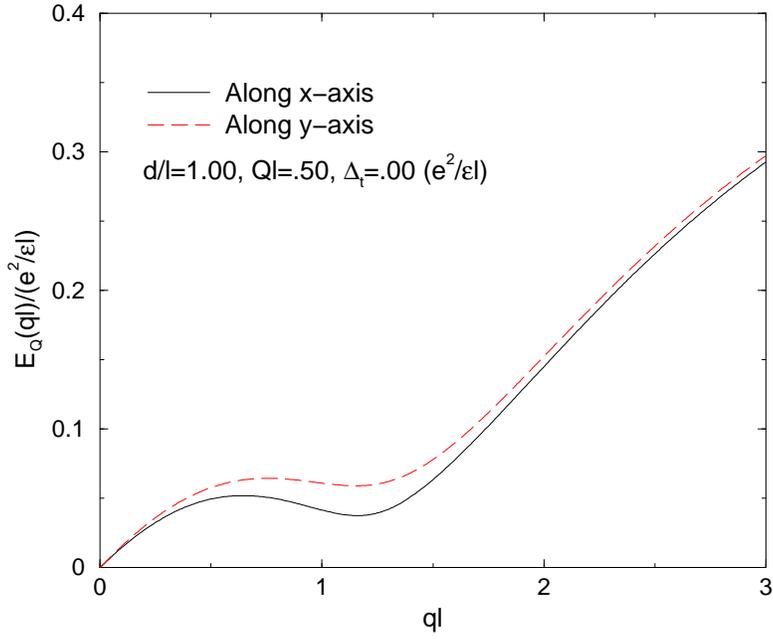}
\vspace{1cm}
\caption{Typical collective mode dispersion for the spiral state. The anisotropy in the dispersion is 
related to the choice of the spiral wavevector $\vec{Q}=Q\hat{x}$. The non-monotonic behavior is 
due to \emph{competing} Hartree and exchange fluctuations present in this system. In our functional integral 
approach, the spin-stiffness can be related to the dependence of the collective mode energies on the spiral 
wavevector.}
\label{fig: dispersion}
\end{center}
\end{figure}

Figure~\ref{fig: dispersion} shows a typical plot of collective-mode energy $E_{sw}(\vec{q},Q)$. We emphasize 
that the anisotropic nature of the dispersion is due to the choice of the spiral wavevector $\vec{Q}=Q\hat{x}$
 and the non-monotonic nature of the energy is due to the competition between the Hartree and exchange terms in 
$a_Q(q)$. As $q\rightarrow\infty$ the 
collective-mode energy approaches the quasiparticle gap $\Delta_Q$. For the uniform case ($Q=0$) the
 dispersion (~\ref{eq: adls14}) reduces to the results previously obtained by diagrammatics~\cite{haf}
 and single-mode approximation~\cite{apb}.

Now let us consider the explicit expressions for contributions to the grand potential due to various
 sets of diagrams mentioned in the previous section. The functional integral 
approach which we have discussed in previous sections gives the following fluctuation contribution to 
the grand potential
\begin{equation}
\label{eq: adls19}
\Omega_c=\frac{1}{2\beta}\ln\prod_{\vec{p},i\Omega_n}
\left[\frac{i\Omega^2_n-E^2_{sw}(\vec{p},Q)}{i\Omega^2_n-\Delta_Q^2}\right]-\frac{1}{2}\mbox{ tr}(V^AD),
\end{equation}
where the first term - $\det\left(1+V^A D\right)$ - has been expressed in terms of the collective-mode 
frequencies $E_{sw}$ and the quasiparticle gap $\Delta_Q$ [see Eq.(~\ref{eq: affia20})]. In contrast, 
the contribution to the grand potential obtained by summing bubbles with vertex corrections - the approximation 
which has been used in the literature~\cite{kmoonprl} - is given by
\begin{equation}
\label{eq: adls20}
\Omega_{bl}=\frac{1}{2\beta}\ln\prod_{\vec{p},i\Omega_n}
\left[\frac{i\Omega^2_n-E^2_{sw}(\vec{p},Q)}{i\Omega^2_n-{\cal E}^2(\vec{p},Q)}\right]-
\frac{1}{2}\mbox{ tr}(V\chi_l).
\end{equation}
It is clear from Eqs.(~\ref{eq: adls19}) and (~\ref{eq: adls20}) that the two approximations are 
different.

\begin{figure}[h]
\begin{center}
\epsfxsize=4in
\epsffile{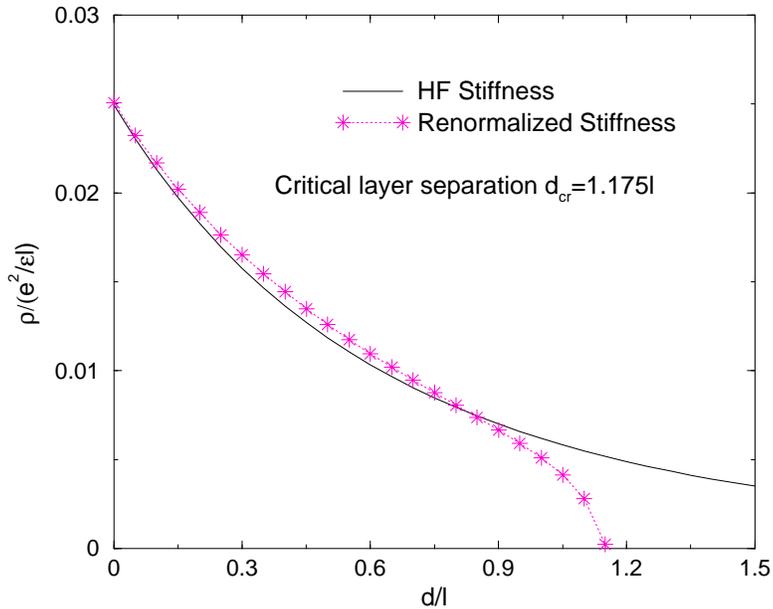}
\vspace{1cm}
\caption{Renormalization of the pseudospin stiffness due to quantum fluctuations. The stiffness is enhanced 
by fluctuations at typical values of $d$. The stiffness vanishes rapidly close to the phase boundary 
$d_{cr}$ as discussed in the text. Note the absence of renormalization at $d=0$, when the Hartree-Fock state 
is the exact ground state. }
\label{fig: stiffness}
\end{center}
\end{figure}

We now calculate the effect of these collective modes on pseudospin stiffness and pseudospin 
polarization of the system. These parameters are obtained from the ground state energy $E_G(Q)$ by 
taking appropriate derivatives
\begin{eqnarray}
\label{eq: adls22}
\rho & = & \frac{1}{A}\frac{\partial^2E_G(Q)}{\partial Q^2}|_{Q=0}, \\
\label{eq: adls23}
m_x  & = & -\frac{4\pi l^2}{A}\frac{\partial E_G(0)}{\partial \Delta_t}.
\end{eqnarray}
In the following we use Eqs.(~\ref{eq: affia17}) and (~\ref{eq: affia19}) to approximate the ground 
state energy. For the spiral-state case, neglecting the kinetic contribution (which is $Q$-independent), 
we obtain
\begin{eqnarray}
\label{eq: adls24}
\frac{E_{qp}}{A} &= &-\frac{1}{4\pi l^2}\left[\frac{\Gamma_A(0)}{2}+\frac{\Delta_Q}{2}\right],\\
\label{eq: adls25}
\frac{E_c}{A} & = & \frac{1}{2A}\sum_{\vec{p}}\left[E_{sw}(\vec{p},Q)-\Delta_Q\right]-
\frac{1}{4\pi l^2}\left[\frac{\Gamma_A(0)}{2}-\frac{\Delta_Q}{2}\right].
\end{eqnarray}
The Hartree-Fock stiffness is obtained from Eq.(~\ref{eq: adls22}) when the ground state energy is 
approximated by $E_{qp}$. The renormalized stiffness is obtained by using $E_G=E_{qp}+E_c$. 
Note that the only $Q$-dependence in $E_G$ is through the term $\left[E_{sw}(\vec{p},Q)-\Delta_Q\right]$. 
Using Eq.(~\ref{eq: adls22}) we obtain the following explicit expression for the renormalized pseudospin 
stiffness, including correlations
\begin{equation}
\label{eq: adls26}
\rho = \frac{1}{2A}\sum_{p}\left[\frac{a''_p b_p+a_pb''_p}{2E_{sw}(p)}+8\pi l^2\rho_{HF}\right].
\end{equation}
This equation is one of the principle results of this work. Here $a_p=a_{Q=0}(p)$ and so on, the primes 
denote second derivative with respect to the wavevector $Q$ and we have used the fact that 
$\Delta''_{Q=0}=-8\pi l^2\rho_{HF}$. 
Figure~\ref{fig: stiffness} compares the Hartree-Fock and the renormalized pseudospin stiffness. At 
$d=0$ the Hartree-Fock ground state is exact and as expected, there is no stiffness renormalization.
 We note that for small layer separations $d\leq l$ fluctuations \emph{enhance} the pseudospin
 stiffness, a possibility that has not been anticipated previously. In our theory $a_p$ vanishes at a 
finite value of $p>l^{-1}$ at layer separations that exceed a critical value $d_{cr}$. In Hartree-Fock 
theory, the ground state changes from a uniform coherent state to a coherent pseudospin-density wave state at 
this point~\cite{cote}. This instability point has previously been identified~\cite{apb}, on the basis of 
heuristic arguments, with a phase transition at which pseudospin ferromagnetism is lost~\cite{sgm}. As $a_p$ 
approaches zero, the first term in Eq.(~\ref{eq: adls26}) becomes large and negative and $\rho$ reaches zero 
for $d$ slightly smaller than $d_{cr}$. While our Gaussian fluctuation calculations are certainly not 
systematic in this regime, our result that the stiffness goes to zero rapidly for $d$ near $d_{cr}$ is 
consistent with numerical exact-diagonalization estimates~\cite{km} and recent suggestions that pseudospin 
ferromagnetism could be lost via a first order transition~\cite{sgm}.

\begin{figure}[h]
\begin{center}
\epsfxsize=4in
\epsffile{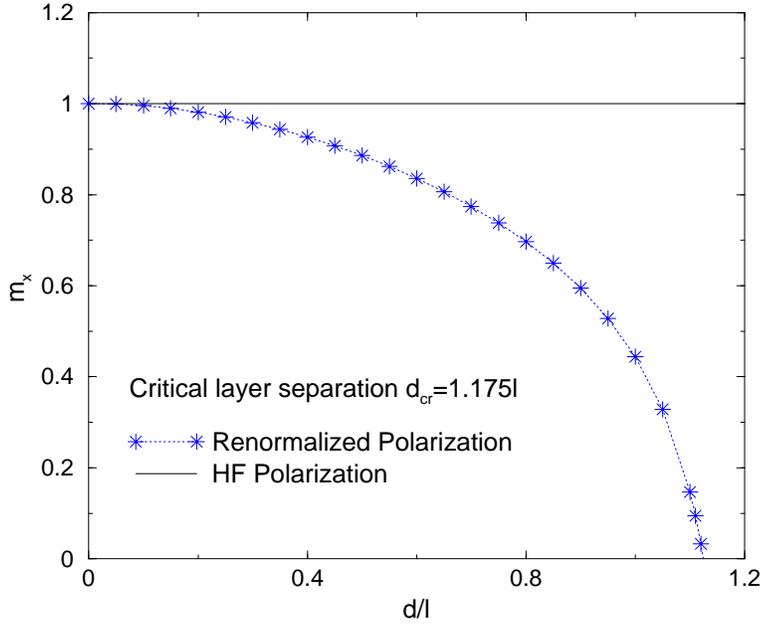}
\vspace{1cm}
\caption{ Dependence of the dimensionless order parameter on layer separation. The mean-field polarization is 
independent of layer separation $d$ and is not sensitive to the changes in collective mode energies that 
occur as $d$ approaches $d_{cr}$. The renormalized polarization vanishes rapidly close to the phase boundary 
$d_{cr}$.}
\label{fig: mx}
\end{center}
\end{figure}

To calculate the order parameter renormalization we start with a system having a finite tunneling 
amplitude $\Delta_t$. In this case the pseudospin symmetry in the $x-y$ plane is explicitly broken, and the 
Hartree-Fock mean-field state is uniformly polarized along the $x$-axis. 
The quasiparticle gap $\Delta_{SAS}$ includes the external tunneling 
contribution $\Delta_t$ and the collective mode is gapped at zero wavevector. Using the finite tunneling and 
zero spiral wavevector version of Eqs.(~\ref{eq: adls24}), (~\ref{eq: adls25}) 
we get the following expression for the renormalized order parameter
\begin{equation}
\label{eq: adls27}
m_x=1-\frac{2\pi l^2}{A}\sum_{\vec{p}}\left[\frac{\epsilon_p-E_p}{E_p}\right],
\end{equation}
where $\epsilon_p=(a_p+b_p)/2$. This analytic expression is the second principle result of this calculation.
 The two terms on the right hand side of Eq.(~\ref{eq: adls27}) have 
a simple interpretation. First term is the Hartree-Fock result for the order parameter ($m_x=1$) and
 the second term represents its suppression due to collective excitations. 
Figure~\ref{fig: mx} compares the Hartree-Fock and renormalized order parameter. When $d=0$, we have
 $\epsilon_p=E_p$ and $m_x=1$. As $d\rightarrow d_{cr}$, collective modes mix more strongly  
into the mean-field ground state and the order parameter is suppressed. We remark that these renormalized 
parameters are qualitatively different from the parameters obtained in the literature~\cite{kmoonprl} by 
using Eq.(~\ref{eq: adls20}) as an approximation for fluctuation correction to the grand 
potential.


\section{Discussion}
\label{sec: discussion}

We have presented a functional integral approach to systematically calculate fluctuation corrections to 
a mean-field approximation for the grand potential. These corrections represent the contribution due to 
collective modes around an ordered mean-field ground state. Our approach takes into account fluctuations 
in both, direct and exchange, particle-hole channels. The same expression for fluctuation corrections to 
the grand potential, Eq.(~\ref{eq: affia18}), can be systematically derived by coupling constant integration from 
GRPA particle-hole response functions. Remarkably, for bilayer quantum Hall systems it is possible to calculate these 
corrections analytically. The pseudospin stiffness and order parameter for the phase-coherent state are 
obtained from ground state energy by taking appropriate derivatives. The two central results of our work are the 
explicit expressions, Eq.(~\ref{eq: adls26}) and Eq.(~\ref{eq: adls27}), for renormalized pseudospin stiffness 
$\rho$ and the renormalized order parameter $m_x$ respectively. We note that these expressions, derived from a 
microscopic treatment, are very similar to those obtained by doing a linearized spin-wave analysis of 
an easy-plane $XY$ spin model on a lattice. In case of a lattice model, though, the two-dimensional momentum 
sum would be cut off at $p=2\pi/a$ where $a$ is the lattice constant.

Our approach presented here is quite generic and applicable to systems where fluctuations in more than one 
channel dominate the low-energy physics. A relevant example is the physics of underdoped cuprates where, 
presumably, fluctuations in the superconducting and antiferromagnetic channels are important to low-energy 
physics. The approach presented here can be used to calculate, for example, the renormalization of superfluid 
density or penetration depth due to quantum fluctuations. 

In this paper, we have presented spiral states as solutions to the mean-field equations when the 
interlayer tunneling is absent, $\Delta_t=0$. In such states, the relative phase between single-particle 
states localized 
in the top and the bottom layer varies as $e^{iQx}$. These spiral states are also realized as mean-field 
ground states in bilayer systems with \emph{nonzero} interlayer tunneling in the presence of an 
\emph{in-plane} field $B_{||}$~\cite{sqm}. In the presence of such a field, the Aharonov-Bohm phase picked up 
by an electron tunneling from top layer to the bottom layer varies precisely in the same manner with 
wavevector $Q=dB_{||}/l^2 B$. 

The results presented here can be easily generalized to include the form-factors arising from finite 
well-widths. It is also possible to take into account the effects of finite temperature by using 
grand potential instead of the ground state energy to calculate the renormalized parameters, though it will 
require some more work. 

In this work we have limited ourselves to Gaussian fluctuations around the Hartree-Fock state. This means 
we are treating the collective modes as non-interacting bosons. If we were to go beyond the Gaussian approximation
 and include higher-order fluctuations, they would generate interactions between these collective modes, which 
are certainly important as the layer separation approaches critical layer separation, $d\rightarrow d_{cr}$. 
A self-consistent treatment of fluctuations is required in this regime. Since we know self-consistent 
approximation schemes for spin-models on a lattice, it is an interesting problem to develop effective 
spin-models which incorporate the microscopic physics of bilayer systems. In the present treatment, the 
strong renormalization of the macroscopic parameters \emph{only} close to the phase-boundary indicates that 
the Gaussian fluctuation approximation is a reasonable approximation over a large range of interlayer 
separation. 

\section{acknowledgements}
The authors acknowledge helpful conversations with Anton Burkov, Steve Girvin, Charles Hanna, Tatsuya 
Nakajima and Jairo Sinova. This work was supported by the Welch Foundation and 
by NSF-DMR0115947.


\end{document}